\documentclass[preprint,preprintnumbers,amsmath,amssymb,floatfix,showpacs,aps]{revtex4}
\usepackage{epsfig}
\usepackage[english]{babel}

\begin{document}
\title{Calculation of intermediate--energy electron--impact ionization of molecular hydrogen and nitrogen using paraxial approximation}
\author{Vladislav V. Serov}
\affiliation{Department of Theoretical Physics, Saratov
State University, 83 Astrakhanskaya, Saratov 410012, Russia}

\date{\today}

\begin{abstract}
We have implemented the paraxial approximation followed by the time-dependent Hartree-Fock method with frozen core for the single impact ionization of atoms and two-atomic molecules. It reduces the original scattering problem to the solution of a five-dimensional time-dependent Schr\"odinger equation. Using this method we calculated the multi-fold differential cross section of the impact single ionization of the helium atom, the hydrogen molecule and the nitrogen molecule by the impact of an intermediate--energy electrons. Our results for the He and the H$_2$ are quite close to the experimental data. Surprisingly, for N$_2$ the agreement is good for the paraxial approximation combined with first Born approximation, but worse for pure paraxial approximation, apparently because of the insufficiency of the frozen core approximation.
\end{abstract}


\maketitle

\section{Introduction}
The electron--impact ionization processes play a crucial role in the physics of the high atmospherical layers, astrophysics, radiation damage of biological objects, and controlled fusion facility operation. That is why these processes remain a subject of researchers interest for long time (see, e.g., \cite{Dunn1985}). Modern research activity in this area is coherent with the progress of experimental instrumentation based on the coincidence technique providing multiple differential cross section (MDCS). The (e,2e) experiments aimed at exploring the ionization dynamics have the experimental geometry different from that of electron momentum spectroscopy \cite{McCarthy1991,Berakdar2001} aimed at getting information about the wave function of the target. MDCS data, for the experimental geometries providing information about dynamics of the ionization process, were collected for (e,2e) on atoms \cite{LahmamBennani1999,Murray1999,Murray2005,Dorn2006,Lohmann2010,Ren2011}, diatomic \cite{LahmamBennani1989,Murray2005H2,LahmamBennani2007,LahmamBennani2009,Chatterjee2009,Senftleben2010} and polyatomic molecules \cite{LahmamBennani2009,LahmamBennani2009b} for the cases of fast \cite{LahmamBennani1999,LahmamBennani1989,Chatterjee2009}, intermediate \cite{Dorn2006,Lohmann2010,Ren2011,LahmamBennani2007,LahmamBennani2009,Senftleben2010,LahmamBennani2009b} and low-energy incident electrons \cite{Murray1999,Murray2005,Murray2005H2,Dorn2006}. For the interpretation of these results, theorists used both perturbative methods, based on the approximate wave functions of electron continuum (see \cite{Popov2010} and references therein), and {\it ab-initio} methods, such as convergent close coupling (CCC) \cite{ReviewBray2002}, time-dependent close coupling (TDCC) \cite{ReviewPindzola2007}, external complex scaling (ECS) \cite{ReviewMcCurdy2004}, and $R$--matrix with pseudostates (RMPS) \cite{Bartschat1996}. Ionization of diatomic molecules is a subject of special interest for both theorists and experimentalists, because this is a natural model for demonstration of Young--type double--slit interference \cite{Chatterjee2009}.
 Recently a large amount of experimental data on ionization of molecular targets by intermediate--energy (hundreds of eV) electrons with small--energy electron ejection appeared \cite{LahmamBennani2007,LahmamBennani2008,LahmamBennani2009,LahmamBennani2009b,Senftleben2010}. These data are hardly interpreted by the theory since the correct (e,2e) description in such circumstances requires correct consideration of higher terms of the Born expansion for scattered electron, the multi--center character of the target, and influence of the residual target electrons on the ejected one. Direct calculations for intermediate--energy electron using ECS or TDCC require grids with small step and, as a consequence, huge computer resources, while the CCC method at present was implemented in molecules only within the single-center approximation  \cite{Kheifets2005}.

In the present work a method is developed that is analogous to the well--known paraxial approximation (PA) in waveguide optics. It allows to reduce the scattering problem solution to the temporal evolution problem. Previously we already implemented a restricted version of this method, i.e. the paraxial approximation combined with the first Born approximation (PA1B), for the calculation of the impact ionization of molecular hydrogen ion H$_2^+$ \cite{Serov2001} and helium atom \cite{Serov2007}. Here we implement the paraxial approximation without Born expansion. To describe the evolution of the multielectron target the time-dependent Hartree--Fock method with frozen core is proposed. As a result, the original problem of the intermediate--energy electron scattering on a multielectron target is reduced to the solution of a five-dimensional time-dependent Schr\"odinger equation. 

The paper is organized as follows. In Section \ref{PAsection} the derivation of the paraxial equation for the scattering problem is briefly described. Section \ref{HFFCAsection} represents the time-dependent Hartree--Fock method with frozen core. In Section \ref{NumMsection} we describe the numerical methods used to solve the obtained five-dimensional time-dependent Schroedinger equation in the cases of one--center and two--center targets. Finally, in Section \ref{PARessection} we present the results of calculating of the multifold differential cross sections of the intermediate--energy electron--impact single ionization for helium atom, non-aligned and aligned hydrogen molecule, and nitrogen molecule in comparison with experimental data and calculations by other authors.

\section{Paraxial approximation}\label{PAsection}
The stationary Schr\"odinger equation describing a projectile with the initial momentum $k_i$ and the coordinate $\mathbf{r}_0$, and a target with a single active electron having the coordinate $\mathbf{r}_1$, has the form (here and below we use atomic units where the Planck's constant  $\hbar$, the absolute value of electron charge $e$ and the electron mass $m_e$ equal unity, $\hbar=|e|=m_e=1$) 
\begin{eqnarray}
\left[-\frac{1}{2\mu}\nabla_0^2-qU_i(\mathbf{r}_0)+\widehat{H}_i(\mathbf{r}_1)-\frac{q}{|\mathbf{r}_1-\mathbf{r}_0|}\right]\Psi(\mathbf{r}_0,\mathbf{r}_1)=\left(\frac{k_i^2}{2\mu}+\epsilon_i\right)\Psi(\mathbf{r}_0,\mathbf{r}_1)
\end{eqnarray}
where $\epsilon_i$ is the initial target energy, $\widehat{H}_i$ is target effective Hamiltonian,  $U_i$ is the effective potential of the residual ion, $\mu$ is a mass of the projectile, and $q$ is a projectile charge.
If we represent the wave function as
\begin{eqnarray}
\Psi(\mathbf{r}_0,\mathbf{r}_1)=\tilde{\Psi}\left(\boldsymbol{\rho}_s,z_0,\mathbf{r}_1\right)\exp\left(ik_iz_0\right),
\end{eqnarray}
where $\boldsymbol{\rho}_s$ is a two--components vector composed from coordinates of $\mathbf{r}_0$ perpendicular to $\mathbf{k}_i$, and neglect the second derivative with respect to $z_0$, then we arrive at the equation akin to the time--dependent one:
\begin{eqnarray}
i\frac{\partial\psi(\mathbf{r}_1,\boldsymbol{\rho}_s,t)}{\partial t}=\left[-\frac{1}{2\mu}\nabla_\perp^2-qU_i(\mathbf{r}_0)+\widehat{H}_i(\mathbf{r}_1)-\frac{q}{|\mathbf{r}_1-\mathbf{r}_0|}\right]\psi(\mathbf{r}_1,\boldsymbol{\rho}_s,t).\label{paraxialeq}
\end{eqnarray}
Here $t=z_0\mu/k_i$ is the time-like parameter and  
$\psi(\mathbf{r}_1,\boldsymbol{\rho}_s,t)=
\tilde{\Psi}\left(\boldsymbol{\rho}_s,k_it/\mu,\mathbf{r}_1\right)\exp(i\epsilon_it)$ is the envelop function. 
The initial condition has the form
\begin{eqnarray}
\psi(\mathbf{r}_1,\boldsymbol{\rho}_s,t_0)=\varphi_i(\mathbf{r}_1)\exp(-i\epsilon_it_0),
\label{InitialCond}
\end{eqnarray}
where $t_0\to -\infty$. The distinction between the paraxial approximation and the well-known eikonal approximation is the fact that the second derivatives with respect to the transverse coordinates of fast particle are not neglected. 

The scattering amplitude can be expressed via the Fourier component with respect to the incoming particle transversal variables (see Eq.(\ref{f_via_psiKperp}) in the Section \ref{NumMsection})
\begin{eqnarray}  
\psi_{\mathbf{K}_{\perp}}(\mathbf{r}_1,t)=\frac{1}{2\pi}e^{i\frac{K_{\perp}^2}{2\mu}t}\int\exp(-i\mathbf{K}_{\perp}\boldsymbol{\rho}_s)\psi(\mathbf{r}_1,\boldsymbol{\rho}_s,t)\,d\boldsymbol{\rho}_s. \label{FouriePsiPA}
\end{eqnarray}
Here in the limit $t\to \infty$ $\mathbf{K}_{\perp}$ is the transverse component of the transferred momentum $\mathbf{K}$, $K_{\perp}=k_s\sin\theta_s$, where $\theta_s$ is the scattering angle. 

 In \cite{Serov2001} a simplified approach, based on the combination of the paraxial approximation with the first Born approximation (PA1B), was proposed. In this approach the scattering problem reduced to the solution of a Schr\"odinger--like inhomogeneous time--dependent equation
\begin{eqnarray}
i\frac{\partial \psi_{\mathbf{K}_{\perp}}(\mathbf{r}_1,t)}{\partial t}&=&
\widehat{H}_i(\mathbf{r}_1)\psi_{\mathbf{K}_{\perp}}(\mathbf{r}_1,t)+F_{\mathbf{K}_{\perp}}(\mathbf{r}_1,t), \label{PA1Beq}
\end{eqnarray}
with the initial condition $\psi_{\mathbf{K}_{\perp}}(\mathbf{r}_1,t_0)=0$. The source term has the form 
\begin{eqnarray}
F_{\mathbf{K}_{\perp}}({\bf r},t)=-\frac{q}{K_{\perp}}
\exp\left[i\left(\frac{K_{\perp}^2}{2}-\epsilon_i\right)t\right]e^{-K_{\perp}|k_i t-z|+i\mathbf{K}_{\perp}\cdot{\mathbf r}_{\perp}}.
\label{MatrixEl}
\end{eqnarray}
This approach is used here for verification and comparison with the pure PA results.

The PA is valid when neglected second derivative of the envelope function $\psi(\mathbf{r}_1,\boldsymbol{\rho}_s,z_0\mu/k_i)$ with respect to $z_0$ is small. It is equivalent to condition \cite{Serov2001}
\begin{eqnarray}
\frac{K_{\perp}^2/2\mu+\Delta\epsilon}{E_i}\ll 1,\label{ValCond}
\end{eqnarray}
where $E_i=k_i^2/2\mu$ is a projectile energy, $\Delta\epsilon$ is a change of the target energy after impact, $\Delta\epsilon=E_e-\epsilon_i$ in the case of the impact ionization, where $E_e$ is a ejected electron energy. Following inequation (\ref{ValCond}), the PA is valid when both angle of scattering $\theta_s\ll 1$ radian, and energy of the ejected electron $E_e\ll E_i$.

\section{Frozen core approximation}\label{HFFCAsection}
Let us start from the time-dependent Hartree--Fock equation for a system contains $N_e$ electrons
\begin{eqnarray}
i\frac{\partial\psi_i(\mathbf{r},t)}{\partial t}=\widehat{F}\left[\{\psi_j(\mathbf{r}',t)\}_{j=1}^{N_o}\right]\psi_i(\mathbf{r},t)+v(\mathbf{r},t)\psi_i(\mathbf{r},t).
\end{eqnarray}
Here the Fock operator for the set of orbital wave functions 
$\{\varphi_j\}_{j=1}^{N_o}$, $N_o=N_e/2$, is
\[
\widehat{F}\left[\{\varphi_j\}_{j=1}^{N_o}\right] =
\widehat{h}+\sum_{j=1}^{N_o}\left(2\widehat{J}[\varphi_j]-\widehat{K}[\varphi_j]\right)
\]
where 
\[
\widehat{h}=-\frac{1}{2}\nabla^2+u(\mathbf{r}),
\]
is the single-electron Hamiltonian, 
\[\widehat{J}[\varphi]\psi(\mathbf{r})=\int\frac{\left|\varphi(\mathbf{r}')\right|^2}{|\mathbf{r}-\mathbf{r}'|}\,d\mathbf{r}'\,\psi(\mathbf{r})\]
is the Coulomb operator, and
\[\widehat{K}[\varphi]\psi(\mathbf{r})=\varphi(\mathbf{r})\int{\frac{\varphi^*(\mathbf{r}')\psi(\mathbf{r}')}{|\mathbf{r}-\mathbf{r}'|}d\mathbf{r}'}\]
is the exchange operator. Let us assume that all orbital wave functions except the $i$--th one are frozen during a process, so that
\[
\psi_j(\mathbf{r},t)=\left\{
\begin{array}{ll}
\psi(\mathbf{r},t),& j=i;\\
\varphi_j(\mathbf{r})\exp(-i\epsilon_j t),& j\neq i;
\end{array}
\right.
\]
where the functions $\{\varphi_j\}_{j=1}^{N_o}$ are solutions of the stationary Hartree--Fock equation
\begin{equation}
\widehat{F}\left[\{\varphi_j\}_{j=1}^{N_o}\right]\varphi_i(\mathbf{r})=\epsilon_i\varphi_i(\mathbf{r}).\label{eigenFock}
\end{equation}

We can introduce an effective potential of the residual molecular ion after the $i$--th electron ejection
\[w_i(\mathbf{r})=2\sum_{j=1}^{N_o}\widehat{J}[\varphi_j]-\widehat{J}[\varphi_i]=\sum_{j=1}^{N_o}(2-\delta_{ij})\int\frac{\left|\varphi_j(\mathbf{r}_2)\right|^2}{r_{12}}\,d\mathbf{r}_2.\]
The residual operator formally has the form
\[\widehat{X}_i=\widehat{F}\left[\{\psi_j\}_{j=1}^{N_o}\right]-\left[\widehat{h}+w_i(\mathbf{r})\right]
=2\widehat{J}[\psi]-\widehat{J}[\varphi_i]-\sum_{j=1}^{N_o}\widehat{K}[\psi_j],\]
or
\[\widehat{X}_i\psi=\left\{\widehat{J}[\psi]-\widehat{J}[\varphi_i]-\sum_{j\neq i}\widehat{K}[\varphi_j]\right\}\psi,\]
but since $\widehat{J}[\psi]$ describes the ejected electron counterpart with a different spin value, whose state can be considered as constant during the process within the frozen core approximation, we get for the exchange operator \[\widehat{X}_i=-\left\{\sum_{j\neq i}\widehat{K}[\varphi_j]\right\}.\]
The correct introduction of this operator gives rise to an integral equation. Since the exchange is essential only in the case when an electron is located near a molecule, we can introduce the approximate exchange operator
\[\widehat{X}_i^{N}=-\widehat{I}^{N}\left\{\sum_{j\neq i}\widehat{K}[\varphi_j]\right\}\widehat{I}^{N}\]
where a projection operator in the subspace of the (\ref{eigenFock}) solutions is
\[\widehat{I}^{N}\psi(\mathbf{r})=\sum_{k=1}^{N}\varphi_k(\mathbf{r})\int\varphi_k^*(\mathbf{r}')\psi(\mathbf{r}')d\mathbf{r}'\]
Then the effective Hamiltonian is
\begin{eqnarray}
\widehat{H}_i=\widehat{h}+w_i(\mathbf{r})+\widehat{X}_i^N \label{effHam}
\end{eqnarray}
If $N\geq N_o$, then the approximate operator provides correct orbital energies for the ground state. In the present work $N=N_o$ was used, which means that we actually neglect the exchange for the continuum states and the excited states. Hence the effective potential of the ion has the form 
\begin{eqnarray}
U_i(\mathbf{r})=u(\mathbf{r})+w_i(\mathbf{r}).
\end{eqnarray}

\section{Numerical method}\label{NumMsection}
The numerical scheme for time propagation is based on the split method. It means that the perpendicular Hamiltonian in Eq.(\ref{paraxialeq})
\begin{eqnarray}
\hat{H}_\perp=-\frac{1}{2\mu}\nabla_\perp^2+U_i(\mathbf{r}_0)+\frac{1}{|\mathbf{r}_1-\mathbf{r}_0|} +\widehat{H}_i(\mathbf{r}_1)
\end{eqnarray}
is splitted to into three parts, for which the time propagation was realized through the Crank-Nicholson method, except the target effective Hamiltonian (\ref{effHam}), for which splitting was performed. The approximation of spatial operators was performed using the discrete variable representation (DVR). 
 
The incoming electron transverse variables were represented in the cylindric coordinate system. For the angular variable $\phi_s$ wave function was expanded in functions
\begin{eqnarray*}
\varphi_{m}(\phi)=\left\{
\begin{array}{ll}
\frac{1}{\sqrt{2\pi}},&m=0;\\
\frac{1}{\sqrt{\pi}}\cos{m\phi},&m>0;\\
\frac{1}{\sqrt{\pi}}\sin{m\phi},&m<0.
\end{array}\right.
\end{eqnarray*}
For the radial variable $\rho_s$ the finite--element method on the Gauss-Lobatto quadratures (FEM--DVR) \cite{Rescigno2000} was used, and the Gauss-Radau quadrature \cite{Tao2009} was used for the first finite element to provide correct boundary conditions at $\rho=0$. After completing the step with the transverse part of the incoming electron kinetic energy operator the discrete Fourier transformation was used to pass to the DVR in the angular variable $\phi_s$ with the quadrature knots
\begin{eqnarray*}
\varphi_j=\frac{2\pi}{N_{\varphi}}(j-1);\;\;\; j=1,\ldots,N_{\varphi},
\end{eqnarray*}
so that the operator of the potential $U_i(\mathbf{r}_0)+1/|\mathbf{r}_1-\mathbf{r}_0|$ became diagonal. But, in order to avoid singularity at $\mathbf{r}_1=\mathbf{r}_0$, the Neumann's expansion \cite{Serov2009}, restricted to $l_{max}=N_{\eta}-1$ and $m_{max}$ (see below), was used for $1/|\mathbf{r}_1-\mathbf{r}_0|$. To provide the second--order precision, the sequence of the split steps is alternated, i.e. the steps were performed as follows: $-\frac{1}{2\mu}\nabla_\perp^2$, the inverse Fourier transformation with respect to $\phi_s$, $U_i(\mathbf{r}_0)+\frac{1}{|\mathbf{r}_1-\mathbf{r}_0|}$, $2\widehat{H}_i(\mathbf{r}_1)$,  $U_i(\mathbf{r}_0)+\frac{1}{|\mathbf{r}_1-\mathbf{r}_0|}$, the direct Fourier transformation, $-\frac{1}{2\mu}\nabla_\perp^2$. Since the temporal grid step was equal to $\tau$, we got the wave function at the time moment $t+2\tau$ after both direct and inverse passing of all split layers.

To perform the split--step procedure with the target Hamiltonian in the case of one-center targets, the method \cite{Melezhik1996} was used, but unlike the authors of \cite{Melezhik1996}, we used the FEM--DVR for the radial variable $r$, and the exterior complex scaling method \cite{Simon1979} was used to suppress non-physical reflection from the boundaries of the $r$--grid. 

Since the wave function converges extremely slowly with the basic functions number growth for the two-center targets with the large nuclear charges, in this case the prolate spheroidal (elliptic) coordinates and the wave function expansion over the basis \cite{Tao2009}
\begin{eqnarray*}
\Phi_{ijm}(\xi,\eta,\phi)=\sqrt{\frac{8}{R^3(\xi_i^2-\eta_j^2)}}f_{mi}(\xi)\varsigma_{mj}(\eta)\varphi_m(\phi).   
\end{eqnarray*} 
were used.
Here $i=1,\ldots,N_r$, $j=1,\ldots,N_{\eta}$, $m=-m_{max},\dots,m_{max}$, $R$ is the distance between the nuclei (in our code, the molecular axis orientation with respect to the incoming direction $\mathbf{k}_i$ could be chosen arbitrary),
\begin{eqnarray*}
f_{mi}(\xi)=\left\{\begin{array}{ll}
f_i(\xi),& \text{ odd } m;\\
\frac{\xi_i}{\sqrt{\xi_i^2-1}}\frac{\sqrt{\xi^2-1}}{\xi}f_i(\xi),& \text{ even } m;
\end{array}\right.        
\end{eqnarray*}
where $f_i(\xi)$ are the FEM--DVR basic functions, composed from pieces of the Lagrange polynomials, meeting the relation $f_i(\xi_{i'})=\delta_{ii'}/\sqrt{w_i}$, $\xi_i$ and $w_i$ are the nodes and weights of a quadrature composed from the Gauss-Radau quadrature for the first finite element and the Gauss-Lobatto quadrature for the rest ones \cite{Tao2009}, and the boundary condition at $\rho=\rho_{max}$ was the Neumann's condition (following Eq.(\ref{InitialCond}), $\lim_{\rho_s\to\infty}\frac{\partial}{\partial \rho_s}\psi(\mathbf{r}_1,\boldsymbol{\rho}_s,t)=0$). The introduction of $f_{mi}(\xi)$ provides correct asymptotic behavior at $\xi=1$ for odd $m$ allowing to save the Legendre function basic feature $f_{mi}(\xi_{i'})=\delta_{ii'}/\sqrt{w_i}$. The angular basis functions were the Legendre functions 
\begin{eqnarray*}
\varsigma_{mj}(\eta)=\sqrt{\varpi_j}\sum_{l=|m|}^{N_{\eta}-1+|m|}\bar{P}^m_l(\eta_j)\bar{P}^m_l(\eta),       
\end{eqnarray*}
where $\eta_j,\varpi_j, j=1,\ldots,N_{\eta}$ are the nodes and weights of the Gauss-Lobatto quadrature in the segment $\eta=[-1,1]$, $\bar{P}^m_l(\eta)$ are the associated Legendre polynomials, orthonormal on the Gauss-Legendre quadrature \cite{Melezhik1996}. Two summands can be distinguished in the Hamiltonian matrix. The first is the quasi--radial Hamiltonian 
\begin{eqnarray*}
H^{mj}_{\xi ii'}=\frac{2}{R^2\sqrt{(\xi_i^2-\eta_j^2)(\xi_{i'}^2-\eta_{j}^2)}}\left[\int_{0}^{\xi_{max}}f_{mi}'(\xi)(\xi^2-1)f_{mi'}'(\xi)d\xi+R(Z_1+Z_2)\xi_i\delta_{ii'}\right],        
\end{eqnarray*}
the matrix having a length halfwith $s$, where $s$ is the finite elements order, and a rank $N_r=sN_{FE}+1$, where $N_{FE}$ is a number of finite elements. The second is the quasi--angular Hamiltonian 
\begin{eqnarray*}
H^{mi}_{\eta jj'}=\frac{2}{R^2\sqrt{(\xi_i^2-\eta_j^2)(\xi_i^2-\eta_{j'}^2)}}\left[\sqrt{\varpi_j\varpi_{j'}}\sum_{l=|m|}^{N_{\eta}-1+|m|}\overline{P}^m_l(\eta_j)l(l+1)\overline{P}^m_l(\eta_{j'})+R(Z_1-Z_2)\eta_j\delta_{jj'}\right]        
\end{eqnarray*}
which is the completely filled square matrix whose rank is $N_{\eta}$. Therefore the Hamiltonian was being splitted into four parts: $\hat{H}_{\xi}$, $\hat{H}_{\eta}$, $U_{sh}(\mathbf{r})=U_i(\mathbf{r})+Z_1/|\mathbf{r}_1-\mathbf{R}/2|+Z_2/|\mathbf{r}_1+\mathbf{R}/2|$ (one contains the average potential of non-active shells and all the nuclei except the first two ones) and $\widehat{X}_i$. After the steps for $\hat{H}_{\xi}$ and $\hat{H}_{\eta}$ had been completed, the transition to DVR was performed for $\phi$ via the Fourier transformation in order to make the matrix of the potential $U_{sh}(\mathbf{r})$ diagonal. The step with the approximate exchange operator $\widehat{X}_i$ was performed as follows. The wave function was expanded by the shell wave functions $\varphi_n(\mathbf{r})$, then the part orthonormal to all them was extracted:
\begin{eqnarray*}
C_n(t)&=&\int\varphi_n^*(\mathbf{r})\psi(\mathbf{r},t)d\mathbf{r};\\
\psi_{rest}(\mathbf{r},t)&=&\psi(\mathbf{r},t)-\sum_{n=1}^{N}C_n(t)\varphi_k(\mathbf{r}).
\end{eqnarray*}
The temporal step for the coefficients $C_n$ was performed using the Crank-Nicholson scheme 
\begin{eqnarray*}
\mathbf{C}(t+\tau)&=&\left[\mathbf{I}+\frac{i\tau}{2}\mathbf{X}\right]^{-1}\left[\mathbf{I}-\frac{i\tau}{2}\mathbf{X}\right]\mathbf{C}(t),
\end{eqnarray*}
where $I_{nk}=\delta_{nk}$, and
\begin{eqnarray}
X_{nk}&=&\langle \varphi_n|\widehat{X}_i|\varphi_k\rangle+\frac{1-\delta_{ni}}{\tau}\delta_{nk}.
\end{eqnarray}
We added large numbers $1/\tau$ to all diagonal elements of the matrix $\mathbf{X}$, except the $i$-th one (it corresponds to the number of the active electron), in order to suppress the active electron transitions to the states occupied by other electrons (this transitions is prohibited by Pauli principle). After performing the step, the wave function part, that had been changed because of the exchange, was added to a residual part 
\begin{eqnarray*}
\psi(\mathbf{r},t+\tau)&=&\psi_{rest}(\mathbf{r},t)+\sum_{n=1}^{N}C_n(t+\tau)\varphi_k(\mathbf{r}).
\end{eqnarray*}
The final steps order in the target electron evolution calculation is: $U_{sh}(\mathbf{r})$, $\widehat{X}_i$, $\hat{H}_{\xi}$, $2\hat{H}_{\eta}$, $\hat{H}_{\xi}$, $\widehat{X}_i$, $U_{sh}(\mathbf{r})$.

The ionization amplitude can be expressed via the Fourier component (\ref{FouriePsiPA}) of the envelop function as \cite{Serov2001}
\begin{equation}
f(\Omega_s,E_e,\Omega_e)=-i k_i \lim_{t\to\infty} \langle \mathbf{k}_e|\psi_{\mathbf{K}_{\perp}}(\mathbf{r},t)\rangle e^{iE_et}\label{f_via_psiKperp}
\end{equation}
Here ${\mathbf k}_e$ is the momentum of the ejected electron, $E_e=k_e^2/2$, $|\mathbf{k}_e\rangle=\varphi_{\mathbf{k}_e}^{(-)}(\mathbf{r})$ are the
continuum wave functions of the target. But we calculated the amplitude using an approach suggested in \cite{Selin1999}, that does not require knowledge of the continuum wave functions in the explicit form. This approach is based on the time Fourier expansion of the probability flux through the boundary 
\begin{eqnarray}
f=-i k_i \int_{t_0}^{T}dt\oint_{S}\mathbf{n}_SdS\cdot
 \mathbf{j}
 \left[\psi_{\mathbf{K}_{\perp}}(\mathbf{r},t),\chi_{\mathbf{k}_e}^{(-)*}(\mathbf{r}) e^{iE_et}\right]. \label{IonizationAmplitude}
\end{eqnarray}
Here the probability flux vector is introduced
\begin{eqnarray}
\mathbf{j}[\psi,\varphi]=\frac{i}{2}[\psi\nabla\varphi-\varphi\nabla\psi],
\end{eqnarray}
$T$ is the time to which evolution is simulated, $S$ is a closed surface around the system (a sphere with the radius $r_S$ in the case of spherical coordinates, or an ellipsoid with the radius $\xi_S=\sqrt{(2r_S/R)^2+1}$ in the case of spheroidal coordinates), $\vec{n}_S$ is its normal vector, $\chi_{\mathbf{k}_e}^{(-)}(\mathbf{r})$ is a function approaching $\varphi_{\mathbf{k}_e}^{(-)}(\mathbf{r})$ at large $r$. In the present work we used an approximated quasiclassical function as $\chi_{\mathbf{k}_e}^{(-)}(\mathbf{r})$, and it differs from the exact continuum function $\varphi_{\mathbf{k}_e}^{(-)}(\mathbf{r})$ by $O(1/r^2)$. 

The problem of this method is that the wave function may not approach zero at the boundary even at a large time value $T$ because the ionization with the ejection of very low energy electrons and transitions to highly excited stationary states is essential. Therefore,
Eq.(\ref{IonizationAmplitude}) yields the value oscillating with growth of $T$. We avoided this artefact by setting 
\begin{eqnarray}
\psi_{\mathbf{K}_{\perp}}(\mathbf{r}_S,t>T)\simeq
\psi_{\mathbf{K}_{\perp}}(\mathbf{r}_S,T)\exp[-iE_{eff}(\mathbf{r}_S,T)(t-T)],
\end{eqnarray}
 where $E_{eff}$ is certain complex ``effective energy'', and then calculating the integral for $t\in (T,\infty)$ analytically. In this case Eq.(\ref{IonizationAmplitude}) turns into
\begin{eqnarray}
f=-i k_i \oint_{S}\vec{n}_S\cdot\left\{\mathbf{j}\left[
\int_{0}^{T}e^{iE_et}\psi_{\mathbf{K}_{\perp}}(\mathbf{r},t)dt-\frac{e^{iE_eT}}{i(E_e-E_{eff})}\psi_{\mathbf{K}_{\perp}}(\mathbf{r},t),
 \chi_{\mathbf{k}_e}^{(-)*}(\mathbf{r}) \right]\right\}dS,
 \label{IonizationAmplitudeCorr}
\end{eqnarray}
where the ``effective energy'' is calculated as 
\begin{eqnarray}
 E_{eff}(\mathbf{r}_S,T)=\frac{i}{\psi_{\mathbf{K}_{\perp}}(\mathbf{r}_S,t)}\frac{\partial \psi_{\mathbf{K}_{\perp}}(\mathbf{r}_S,t)}{\partial t}.
\end{eqnarray}
The validity condition of this approximation $\left|dE_{eff}/dT\right|/E_e^2\ll 1$ at $T\to\infty$ proceed to $[U(r_S)/E_e]^2\sim 1/r_S^2\ll 1$ that coincide with order of the accuracy of   $\chi_{\mathbf{k}_e}^{(-)}(\mathbf{r})$.

\section{Results}\label{PARessection}
\begin{figure}[h!]
\includegraphics[angle=-90,width=0.6\textwidth]{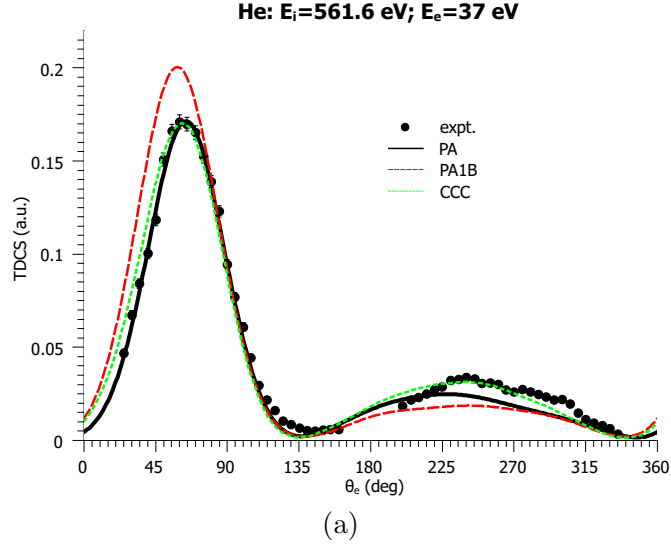}\\
(a)\\
\includegraphics[angle=-90,width=0.6\textwidth]{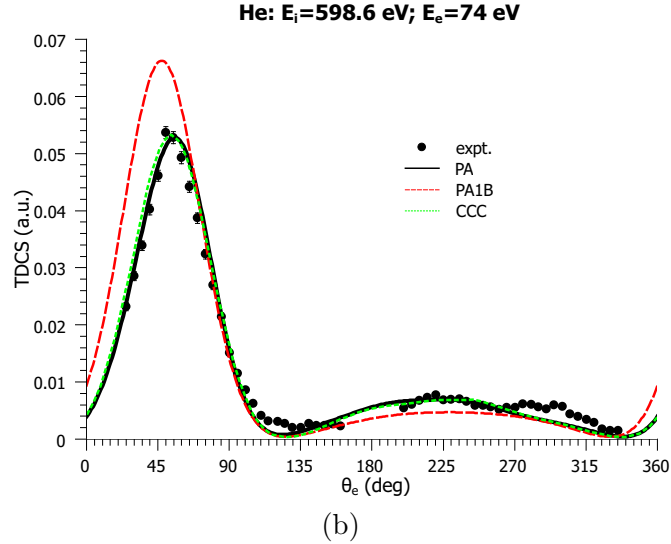}\\
(b)\\
\caption{(Color online) The TDCS of He$(e,2e)$ process as a function of the ejection angle $\theta_{e}$ for ejection energies a) $E_e$=37 eV; b) $E_e$=74 eV: our PA results (solid line), our PA1B results (dashed line), CCC results \cite{LahmamBennani2008} (dotted line) and experimental data \cite{LahmamBennani2008} (circles).\label{FIGsigmaHe}}
\end{figure}
To test the method, we calculated the TDCS of the helium single ionization by fast electron impact at the experimental parameters set \cite{LahmamBennani2008}: the scattered electron energy $E_i=500$ eV, the ejected electron energy $E_e=37$ eV and $E_e=74$ eV, in the third experimental data set \cite{LahmamBennani2008} the ejection energy $E_e=205$ is too large for the PA calculation. In  Fig.\ref{FIGsigmaHe} one can see our PA and PA1B results, the experimental data \cite{LahmamBennani2008} and the CCC results from \cite{LahmamBennani2008}. The latter ones are normalized to provide the best coincidence with a binary peak in PA results, since in \cite{LahmamBennani2008} they are given in arbitrary units. It is seen that our PA results coincide very well both with the experiment and with the CCC data, though CCC reproduces the recoil peak at $E_e=37$ eV better, and at $E_e=74$ eV our results are indistinguishable from the CCC results. 

\begin{figure}[h!]
\includegraphics[angle=-90,width=0.6\textwidth]{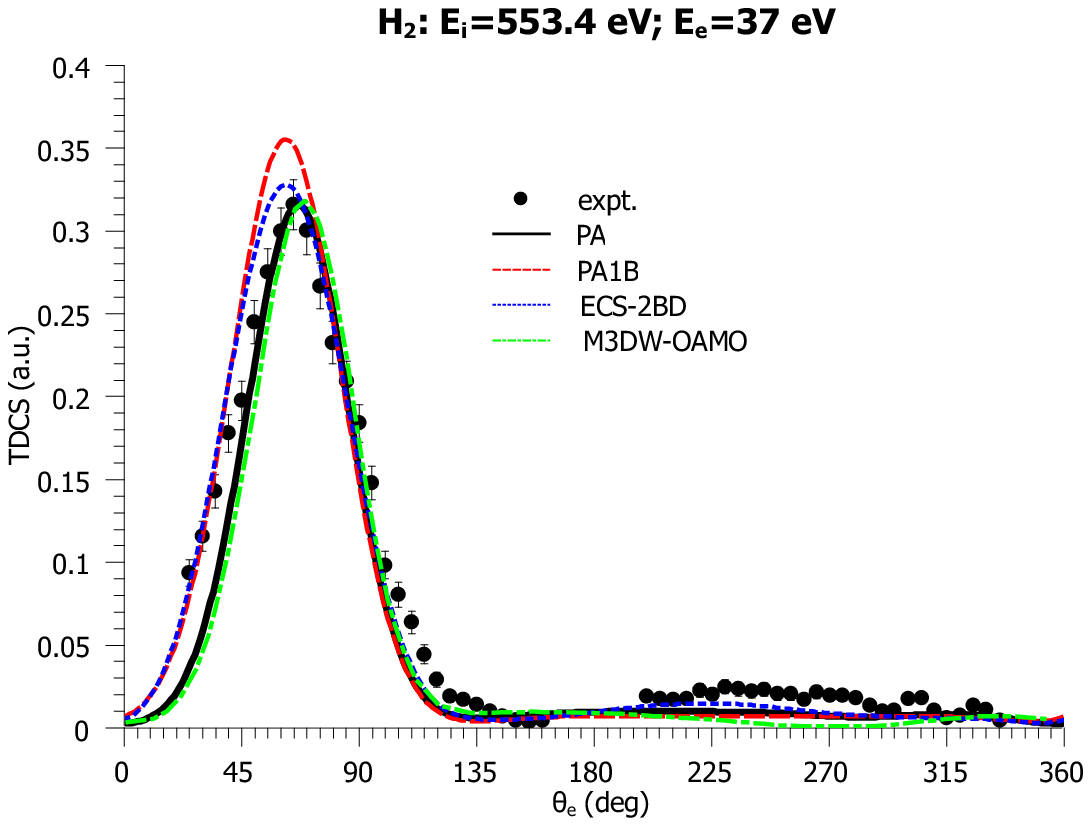}\\
(a)\\
\includegraphics[angle=-90,width=0.6\textwidth]{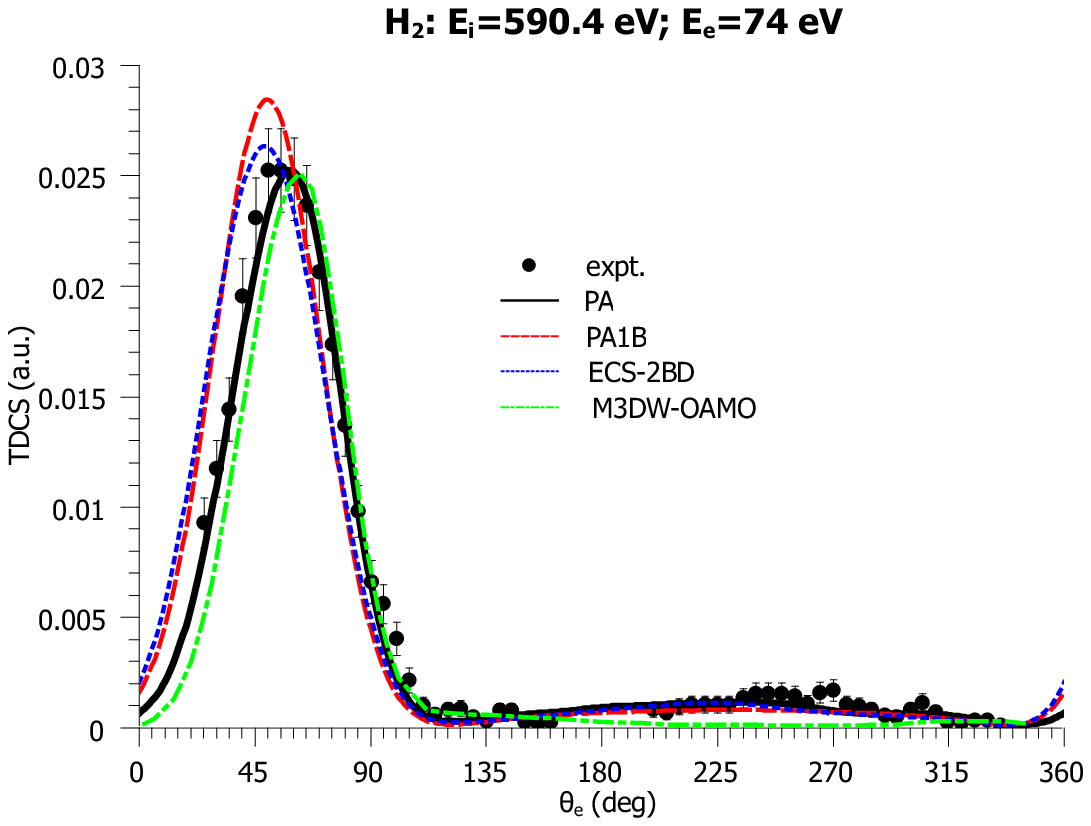}\\
(b)\\
\caption{(Color online) The TDCS of H$_2(e,2e)$ process as a function of the ejection angle $\theta_{e}$ for ejection energies a) $E_e$=37 eV; b) $E_e$=74 eV: our PA results (solid line), our PA1B results (dashed line), ECS-2BD results (dotted line), M3DW-OAMO results \cite{LahmamBennani2008} (dot-dashed line) and experimental data \cite{LahmamBennani2008} (circles).\label{FIGsigmaH2}}
\end{figure}
Further we calculated the TDCS of the single ionization of a non-aligned H$_2$ molecule by the fast electron impact also at the experimental parameters \cite{LahmamBennani2008}. In  Fig.\ref{FIGsigmaH2}, beside the PA and PA1B results and the experimental data \cite{LahmamBennani2008}, we present the results of the external complex scaling method with account second term in Born series in dipole approximation (ECS-2BD) \cite{Serov2010}, and the molecular
three-body distorted wave coupled with an orientation-averaged molecular orbital
approximation (M3DW-OAMO) results \cite{LahmamBennani2008}. The experimental data and the M3DW-OAMO results are normalized to provide the best coincidence with the binary peak in PA results. Our PA results coincide with the experimental ones better than those of M3DW-OAMO both in the binary peaks position and in the recoil peaks magnitude, though the recoil peak is slightly underestimated in our results at $E_e=37$ eV. The ECS-2BD results \cite{Serov2010} coincide well in magnitude with the PA results, but demonstrate strong underestimation of the angular shift with respect to the direction of the momentum transfer vector $\mathbf{K}$. The ECS-2BD method takes into consideration only the second--order Born term bi--dipole component, i.e. the contribution to the second--order Born term that is linear with respect to both radius--vectors of two target electrons. The Hartee--Fock approximation is considered here taking into account interaction of only one target electron with scattered electron but exactly. Hence one can conclude that the main contribution to the angular shift is given by the second Born term components which depend only on the coordinates of one electron. This can explain the failure of the ECS-2BD in describing of the angular distribution of ejected electrons in the H$_2$ double ionization \cite{Serov2010}.    

\begin{figure}[h!]
\includegraphics[angle=-90,width=0.6\textwidth]{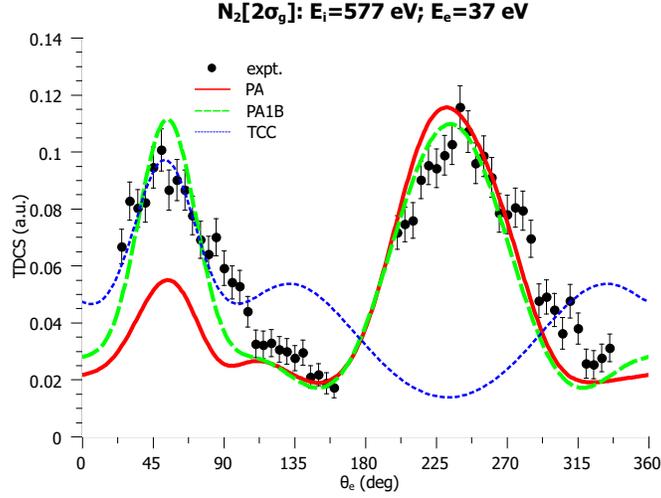}\\
(a)\\
\includegraphics[angle=-90,width=0.6\textwidth]{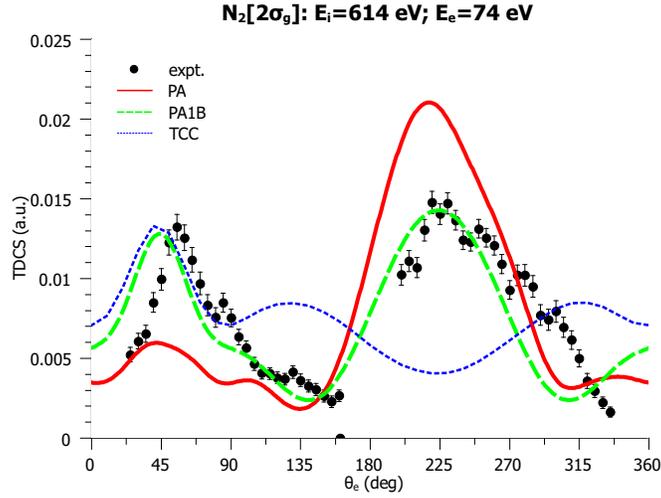}\\
(b)\\
\caption{(Color online) The TDCS of $(e,2e)$ process on $2\sigma_g$ shell of N$_2$ as a function of the ejection angle $\theta_{e}$ for ejection energies a) $E_e$=37 eV; b) $E_e$=74 eV: our PA results (solid line), our PA1B results (dashed line), TCC-1B results (dotted line) and experimental data (circles).\label{FIGsigma2sg}}
\end{figure}
We calculated the TDCS for a non-aligned N$_2$ molecule at the experimental parameters \cite{LahmamBennani2007,LahmamBennani2009}. The variational functions \cite{Scherr1955} were used as the functions of initial states of N$_2$ orbitals. In Fig.\ref{FIGsigma2sg} we show the PA and PA1B results together with the experimental data \cite{LahmamBennani2009}, and also the TCC-1B results \cite{LahmamBennani2008} for the ionization of an electron from an inner $2\sigma_g$ shell of N$_2$. Since the PA1B result is much closer to the experimental data than that of PA, the experimental data and the TCC-1B results are normalized to fit the PA1B. We suppose that such a discouraging PA failure is caused by using the approximation of one active electron and, as a consequence, with neglecting the target interelectron correlation and the change of the rest electrons state during the interaction with the incoming electron.  In the first Born model, the target electron acquires the escape velocity after a single act of momentum transfer from the incident electron and rapidly leaves the molecule, so, other electrons have no time to change their state and the frozen core approximation is correct. High-order Born terms include two-step processes: after the first momentum transfer the electron can stay in the excited state (note, that it is most probable for initially strongly bound internal electrons) and leave the molecule only after the second momentum transfer. Certainly, during the first stage the other electron has time to change its state. Since the interelectron correlation leads to the growth of the average interelectron distance, it should obviously lead to a decrease of the repulsion of the ejected electron from the other target electrons in comparison with the frozen core approximation. So, in the intermediate state of a two-step process, the active electron is actually stronger bound with the molecule then in the frozen core approximation and has less chance to leave the molecule after the second impact. As  result, the ionization via two-step processes is strongly overestimated when using the frozen core approximation. In PA1B the two-step processes are omitted, while in PA with frozen core they are present and strongly overestimated. Apparently, that is why the PA1B turned out to be much closer to the experimental data than the PA. 

\begin{figure}[h!]
\includegraphics[angle=-90,width=0.6\textwidth]{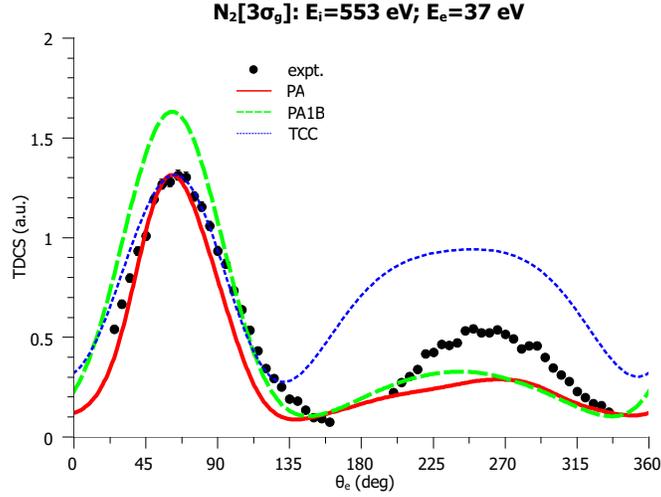}\\
(a)\\
\includegraphics[angle=-90,width=0.6\textwidth]{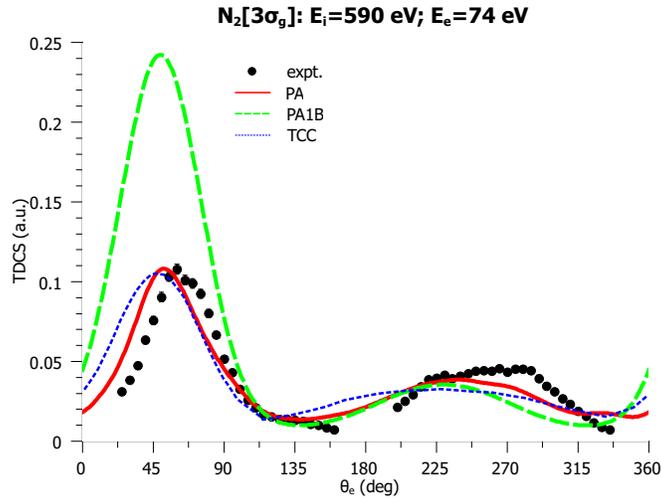}\\
(b)\\
\caption{(Color online) The same as in the Fig.\ref{FIGsigma2sg}, but for the ionization from outer shells. \label{FIGsigmaSum}}
\end{figure}
In Fig.\ref{FIGsigmaSum} the same as in Fig.\ref{FIGsigma2sg} is shown, but for the outer shells ionization. We calculated the contributions to the TDCS given by the ionization from $3\sigma_g$, $1\pi_u$ and $2\sigma_u$--shells of N$_2$, and summarized them with the coefficients 1, 0.78, and 0.32 respectively, following \cite{LahmamBennani2009}. Here the experimental data and the TCC-1B data are rescaled to fit the PA binary peak value. In this case our PA results are obviously closer to the experimental points than those of the PA1B and the TCC-1B, though the PA notably underestimates the recoil peak value at $E_e=37$ eV and the angular shift of the binary peak at $E_e=74$ eV. These distinctions seem to be associated with the target outer shells dynamics. 

\begin{figure}[h!]
\includegraphics[angle=-90,width=0.49\textwidth]{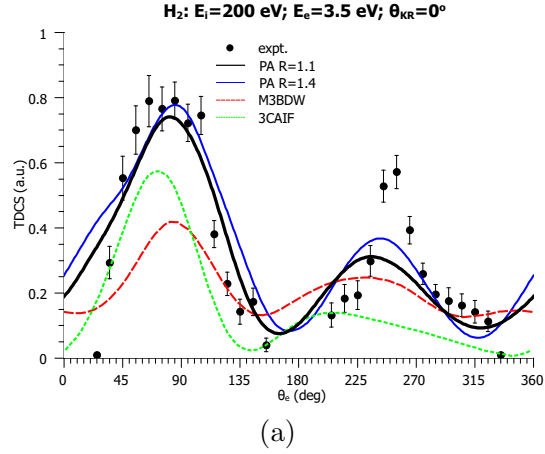}\\
(a)\\
\includegraphics[angle=-90,width=0.49\textwidth]{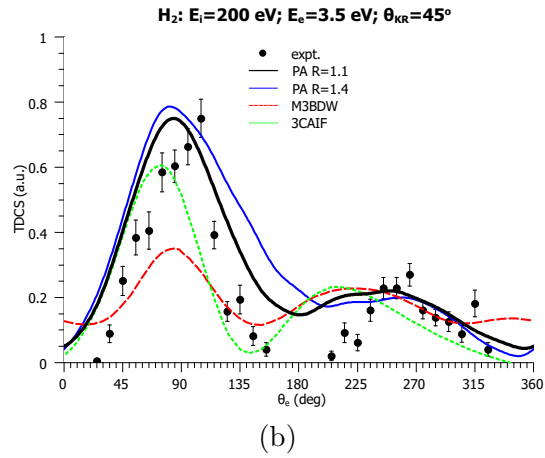}\\
(b)\\
\includegraphics[angle=-90,width=0.49\textwidth]{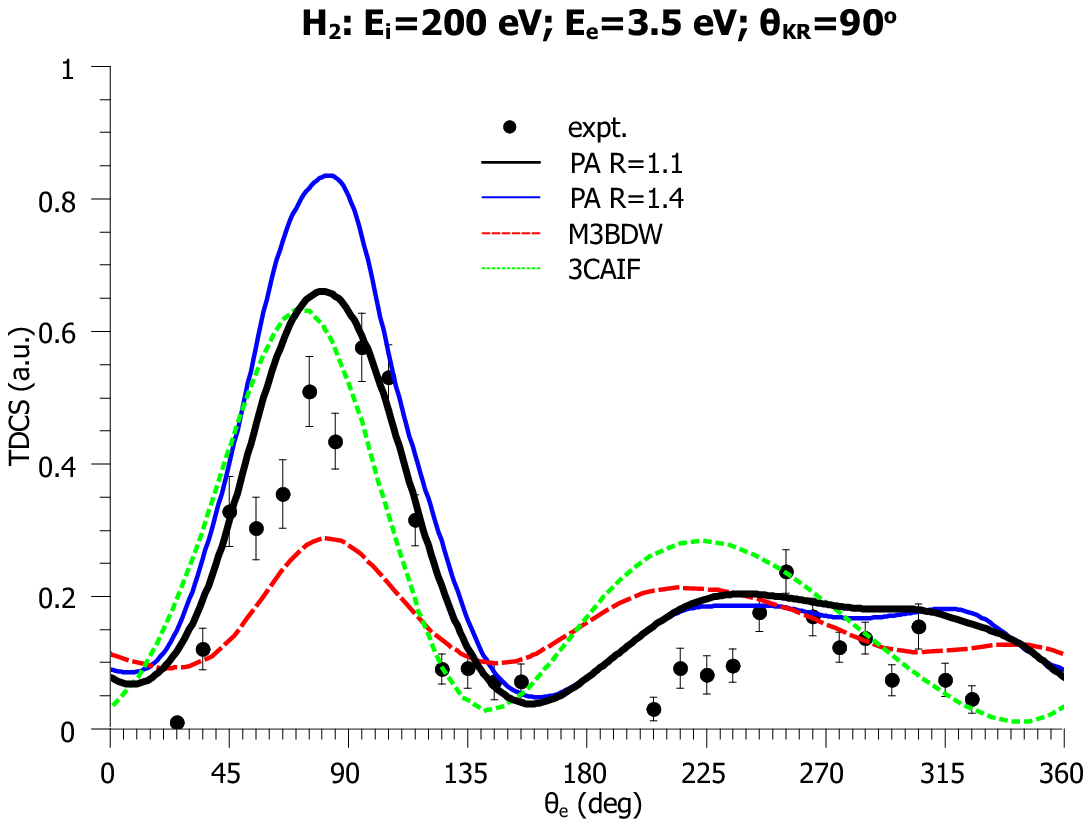}\\
(c)\\
\caption{(Color online) The TDCS of $(e,2e)$ process for an aligned H$_2$ as a function of the ejection angle $\theta_{e}$ for the angle between molecular axis and the momentum transfer direction a) $\theta_{KR}$=0$^\circ$; b) $\theta_{KR}$=45$^\circ$; b) $\theta_{KR}$=90$^\circ$: our PA results for interatomic distance $R=1.1$ (thick solid line) and $R=1.4$ (thin solid line), the M3BDW results (dashed line), the 3CAIF results  (dotted line) and experimental data \cite{Senftleben2010} (circles).\label{FIGsigmaH2aligned}}
\end{figure}
Also we calculated the TDCS of the ionization of aligned H$_2$ for the following experimental parameters \cite{Senftleben2010}: the energy of impact electron $E_i=200$ eV, the scattering angle $\theta_s=16^\circ$, and the ejected electron energy $E_e=3.5$ eV (Fig. \ref{FIGsigmaH2aligned}). Since the incoming electron momentum $k_i=3.83$ is rather small and the scattering angle is large, it can be considered as a rigorous test of the paraxial approximation applicability limits. The initial orientation of the molecule in \cite{Senftleben2010} was measured by registering the protons that appear due to the dissociation of the residual H$_2^+$ ion. Since the dominant channel of this process is ground-state dissociation, we used the internuclear distance $R=1.1$ a.u. for H$_2$, when this process is possible. Fig. \ref{FIGsigmaH2aligned}, beside our PA results and the experimental data \cite{Senftleben2010}, also demonstrates the molecular three-body distorted wave (M3BDW) results and the three-Coulomb wavefunction approach for the helium target multiplied by the interference factor (3CAIF) results from \cite{Senftleben2010}. The experimental data are normalized to binary peak magnitude in our PA $R=1.1$ curve for $\theta_{KR}$=45$^\circ$. It is seen that our method yields the values of binary to recoil peak magnitudes ratio closest to the experimental one compared with other theoretical methods. The distinctions from the experiment may be a consequence of neglecting the ionization--dissociation via autoionizing states in our method, whereas this process gives a significant contribution to the ionization dissociation of the H$_2$ molecule.

\section{Conclusion}
We have calculated the TDCS of the intermediate--energy electron ionization of the helium atom, the hydrogen molecule, and the nitrogen molecule using the method based on the paraxial approximation for the incoming electron and the time-dependent Hartree-Fock method with frozen core for the target electrons. The comparison with experimental data shows that this method works very well both for helium and H$_2$, and quite well for N$_2$ in the case of electron ejection from outer shells. For  inner $2\sigma_g$ shell of N$_2$ the agreement is good for the PA1B, but worse for pure PA, apparently because of the insufficiency of the frozen core approximation. In contrast to original scattering problem, PA can be easy combined, for instance, with time-dependent density functional method allows to calculate a evolution of all electronic orbitals with an approximate account of a interelectronic interaction with an acceptable cost of computer resources. So, we plan to develop PA approach on this way. 
We also plan to use the method (in its present form) to study the dependence of the MDCS on the projectile charge sign. The PA for positron projectile would give results different from those for electron projectile, in contrast with the PA1B.

\acknowledgments The author would like to acknowledge Prof. B. Joulakian for drawing my attention to this area, Prof. A. Lahmam-Bennani and Dr. A. Senftleben for the experimental data sharing, Tatyana Sergeeva and Prof. V. Derbov for help. This work was supported by the President of Russian Federation, grant No. MK-2344.2010.2, and by the Russian Foundation for Basic Research, grant No. 11-01-00523-a.


\begin{thebibliography}{99}
\bibitem{Dunn1985} T. D. M\"ark and G.H. Dunn (Eds.), {\it Electron Impact Ionization} (Springer Verlag, Viena, 1985).
\bibitem{McCarthy1991} I. E. McCarthy and E. Weigold, Rep. Prog. Phys. \textbf{54}, 789 (1991).
\bibitem{Berakdar2001} J. Berakdar and J. Kirschner (Eds.), {\it Many-particle spectroscopy of atoms molecules, clusters, and surfaces} (Kluver Academic/Plenum Publishers, New York, 2001).
\bibitem{LahmamBennani1999} A. Kheifets, I. Bray, A. Lahmam-Bennani, A. Duguet, and I.
Taouil, J. Phys. B \textbf{32}, 5047 (1999). 
\bibitem{Murray1999} N. J. Bowring, F. H. Read, and A. J. Murray, J. Phys. B \textbf{32}, L57 (1999). 
\bibitem{Murray2005} A. J. Murray, Phys. Rev. A \textbf{72}, 062711 (2005). 
\bibitem{Dorn2006} M. D\"urr, C. Dimopoulou, B. Najjari, A. Dorn, and J. Ullrich,
Phys. Rev. Lett. \textbf{96}, 243202 (2006).
\bibitem{Lohmann2010} L. R. Hargreaves, M. A. Stevenson, and B. Lohmann, J. Phys. B \textbf{43}, 205202 (2010).
\bibitem{Ren2011} X. Ren, A. Senftleben, T. Pfl\"uger, A. Dorn, K. Bartschat, and J. Ullrich,
Phys. Rev. A \textbf{83}, 052714 (2011).
\bibitem{LahmamBennani1989} A. Lahmam--Bennani, C. Dupre, and A. Duguet, Phys. Rev. Lett. \textbf{63}, 1582 (1989).
\bibitem{Chatterjee2009} S. Chatterjee, S. Kasthurirangan, A. H. Kelkar, C. R. Stia, O. A. Foj\'on, R. D. Rivarola, and L. C. Tribedi,  J. Phys. B \textbf{42}, 065201 (2009).
\bibitem{Murray2005H2} A. J. Murray, J. Phys. B \textbf{38}, 1999 (2005).
\bibitem{LahmamBennani2007} A. Naja, E.M. Staicu-Casagrande, A. Lahmam-Bennani, M. Nekkab,
F. Mezdari, B. Joulakian, O. Chuluunbaatar, and D. H. Madison, J. Phys. B \textbf{40}, 3775 (2007).
\bibitem{LahmamBennani2009} A. Lahmam-Bennani, E.M. Staicu-Casagrande, and A. Naja, J. Phys. B \textbf{42}, 235205 (2009).
\bibitem{Senftleben2010} A. Senftleben, T. Pfl\"uger, X. Ren, O. Al-Hagan, B. Najjari, D. Madison,
A. Dorn and J. Ullrich, J. Phys. B \textbf{43}, 081002 (2010).
\bibitem{LahmamBennani2009b} A Lahmam-Bennani, A. Naja, E. M. Staicu-Casagrande,
N. Okumus, C. Dal Cappello, I. Charpentier, and S. Houamer, J. Phys. B \textbf{42}, 165201 (2009).
\bibitem{Popov2010} Yu. V. Popov, O. Chuluunbaatar, V. L. Shablov, and K. A. Kouzakov, Physics of Particles and Nuclei \textbf{41}, 543 (2010).
\bibitem{ReviewBray2002} I. Bray, D. V. Fursa, A. S. Kheifets, and A. T. Stelbovics, J. Phys. B \textbf{35}, R117 (2002).
\bibitem{ReviewPindzola2007} M. S. Pindzola \textit{et al.}, J. Phys. B \textbf{40}, R39 (2007).
\bibitem{ReviewMcCurdy2004} C. W. McCurdy, M. Baertschy, and T. N. Rescigno, J. Phys. B \textbf{37}, R137 (2004).
\bibitem{Bartschat1996} K. Bartschat, E. T. Hudson, M.P. Scott, P.G. Burke, and V.M. Burke, J. Phys. B \textbf{29}, 115 (1996).
\bibitem{LahmamBennani2008} E.M. Staicu-Casagrande, A. Naja, A. Lahmam-Bennani, A.S. Kheifets,
 D.H. Madison, and B. Joulakian, J. Phys.: Conf. Ser. \textbf{141}, 012016 (2008).
\bibitem{Kheifets2005} A. S. Kheifets and I. Bray, Phys. Rev. A \textbf{72}, 022703 (2005).
\bibitem{Serov2001} V. V. Serov, V. L. Derbov, B. B. Joulakian, and S. I. Vinitsky,
Phys. Rev. A \textbf{63}, 062711 (2001).
\bibitem{Serov2007} V. V. Serov, V. L. Derbov, B. B. Joulakian, and S. I. Vinitsky,
Phys. Rev. A \textbf{75}, 012715 (2007).
\bibitem{Rescigno2000} T. N. Rescigno and C. W. McCurdy, Phys. Rev. A \textbf{62}, 032706 (2000).
\bibitem{Tao2009} Liang Tao, C. W. McCurdy, and T. N. Rescigno, Phys. Rev. A \text{79}, 012719 (2009)
\bibitem{Serov2009} V. V. Serov and B. B. Joulakian, Phys. Rev. A \textbf{80}, 062713 (2009).
\bibitem{Melezhik1996} V.S. Melezhik, Hyperfine Interactions \textbf{101/102}, 365 (1996).
\bibitem{Simon1979} B. Simon, Phys. Lett. A \textbf{71}, 211 (1979).
\bibitem{Selin1999} A.M. Ermolaev, I.V. Puzynin, A.V. Selin, and S.I. Vinitsky, Phys. Rev. A \textbf{60}, 4831 (1999).
\bibitem{Serov2010} V. V. Serov and B. B. Joulakian, Phys. Rev. A \textbf{82}, 022705 (2010). 
\bibitem{Scherr1955} C. W. Scherr, J. Chem. Phys \textbf{23}, 569 (1955). 
\end{thebibliography}
\end{document}